\documentclass[12pt]{article}
\usepackage{scicite}
\usepackage{times}
\usepackage{graphicx}% Include figure files
\usepackage{dcolumn}% Align table columns on decimal point
\usepackage{bm}% bold math
\usepackage{eucal}

\newenvironment{sciabstract}{%
\begin{quote}}
{\end{quote}}

\newcounter{lastnote}
\newenvironment{scilastnote}{%
\setcounter{lastnote}{\value{enumiv}}%
\addtocounter{lastnote}{+1}%
\begin{list}%
{\arabic{lastnote}.}
{\setlength{\leftmargin}{.22in}}
{\setlength{\labelsep}{.5em}}}
{\end{list}}

\title{Quantum Phase Transition of a Magnet in a Spin Bath}

\author{H. M. R\o{}nnow,$^{1,2,6\ast}$ R. Parthasarathy,$^{2}$ J. Jensen$^{3}$,\\ G. Aeppli$^{4}$, T. F. Rosenbaum$^{2}$, D. F. McMorrow$^{4-6}$\\
\\
\normalsize{$^{1}$Laboratory for Neutron Scattering, ETH-Z\"{u}rich and Paul Scherrer Institut,}\\
\normalsize{5232 Villigen, Switzerland}\\
\normalsize{$^{2}$The James Franck Institute and Department of Physics, The University of Chicago,}\\
\normalsize{IL 60637, USA}\\
\normalsize{$^{3}$\O{}rsted Laboratory, Niels Bohr Institute fAPG,}\\
\normalsize{Universitetsparken 5, 2100 Copenhagen, Denmark}\\
\normalsize{$^{4}$London Centre for Nanotechnology and Department of Physics and Astronomy,}\\ \normalsize{UCL, London, WC1E 6BT}\\
\normalsize{$^{5}$ISIS, RAL, UK}\\
\normalsize{$^{6}$Ris\o{} National Laboratory, DK-4000 Roskilde, Denmark}\\
\normalsize{$^\ast$To whom correspondence should be addressed;
E-mail:  henrik.ronnow@psi.ch.} }

\date{}

\begin{document}

\baselineskip24pt

\maketitle

\begin{sciabstract}
The excitation spectrum of a model magnetic system, LiHoF$_4$, has
been studied using neutron spectroscopy as the system is tuned to
its quantum critical point by an applied magnetic field. The
electronic mode softening expected for a quantum phase transition
is forestalled by hyperfine coupling to the nuclear spins. We show
that interactions with the nuclear spin bath control the length
scale over which the excitations can be entangled. This generic
result limits how far it is possible to approach intrinsic
electronic quantum criticality.
\end{sciabstract}

The preparation and preservation of entangled quantum states is
particularly relevant for the development of quantum computers,
where interacting qubits must produce states sufficiently
long-lived for meaningful manipulation. The state lifetime,
typically referred to as a decoherence time, is derived from
coupling to the background environment. For solid state quantum
computing schemes, the qubits are typically electron spins, and
they couple to two generic background environments\cite{stamp1}.
The oscillator-bath
--- \emph{i.e.} delocalised environmental modes\cite{feynman}, such as thermal
vibrations coupled via magneto-elastic terms to the spins --- can
be escaped by lowering the temperature to a point where the
lattice is essentially frozen. Coupling to local degrees of
freedom, such as nuclear magnetic moments which form a spin-bath,
may prove more difficult to avoid, as all spin-based candidate
materials for quantum computation have at least one naturally
occurring isotope that carries nuclear spin. Experimental work on
this subject has been largely restricted to the relaxation of
single, weakly interacting, magnetic moments such as those on
large molecules\cite{wernsdorfer02} while much less is known about
spins as they might interact in a real quantum computer. In this
regard, the insight that quantum phase
transitions\cite{osterloh02} are a good arena for looking at
fundamental quantum properties of strongly interacting spins turns
out to be valuable, as it has already been for explorations of
entanglement. In particular, we show that coupling to a nuclear
spin bath limits the distance over which quantum mechanical mixing
affects the electron spin dynamics.

Quantum phase transitions (QPTs) are transitions between different
ground states driven not by thermal fluctuations, but by quantum
fluctuations controlled by some parameter such as doping, pressure
and magnetic field\cite{sachdev,sachdevbook}. Much of the interest
in QPTs stems from their importance for understanding materials
with unconventional properties -- heavy Fermion systems and high
temperature superconductors. However, these materials are rather
complex, and do not easily lend themselves to a universal
understanding of QPTs. To this end it is desirable to identify
quantum critical systems with a well-defined and solvable
Hamiltonian, and with a precisely controllable tuning parameter.
One very simple model displaying a QPT is the Ising ferromagnet in
a transverse magnetic field
\cite{degennes,elliott,stinchcombe,sachdev}, with the Hamiltonian:
\begin{equation}
{\mathcal H}=-\sum_{ij}
{\mathcal J}_{ij}\,\sigma_i^z\cdot\sigma_j^z
-\Gamma\sum_i\sigma_i^x
\end{equation}
In the absence of a magnetic field, the ensemble of the two
degenerate $\sigma^z=\pm1$ states orders ferromagnetically below a
critical temperature $T_c$. The transverse field $\Gamma$ mixes
the two states and even at zero temperature leads to destruction
of long-range order in a QPT at a critical field $\Gamma_c$. In
the ferromagnetic state at zero field and temperature, the
excitation spectrum is momentum independent and centered at the
energy $4\sum_j{\mathcal J}_{ij}$ associated with single spin
reversal. Upon application of a magnetic field, however, the
excitations acquire a dispersion, softening to zero at the zone
centre $q=0$ when the QPT is reached.

\begin{figure}[b]
\includegraphics[width=0.8\linewidth]{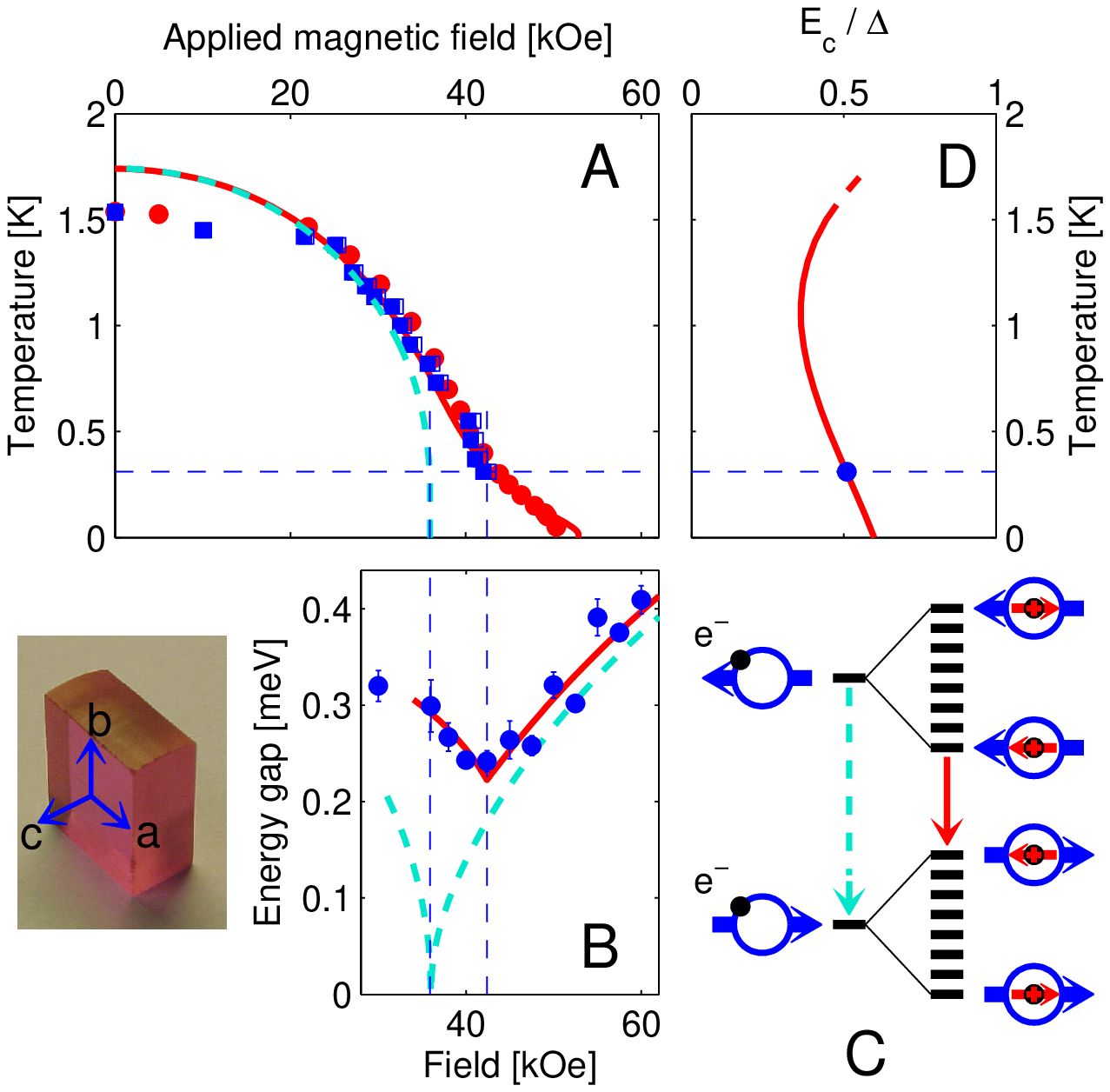}
\caption{ {\bf A} Phase diagram of LiHoF$_4$ as a function of
transverse magnetic field and temperature from susceptibility
\cite{Bitko} (circles) and neutron scattering (squares)
measurements. Lines are $1/z$ calculations with (solid) and
without (dashed) hyperfine interaction. Horizontal dashed guide
marks the temperature 0.31~K at which inelastic neutron
measurements were performed. {\bf B} Field dependence of the
lowest excitation energy in LiHoF$_4$ measured at
$Q=(1+\epsilon,0,1)$. Lines are calculated energies scaled by
$Z=1.15$ respectively with (solid) and without (dashed) hyperfine
coupling. The dashed vertical guides show how in either case the
minimum energy occurs at the field of the transition (c.f. {\bf
A}). {\bf C} Schematic of electronic (blue) and nuclear (red)
levels as the transverse field is lowered towards the QCP.
Neglecting the nuclear spins, the electronic transition (light
blue arrow) would soften all the way to zero energy. Hyperfine
coupling creates a non-degenerate multiplet around each electronic
state. The QCP now occurs when the excited state multiplet through
level repulsion squeezes the collective mode of the ground state
multiplet to zero energy, hence forestalling complete softening of
the electronic mode. Of course, the true ground and excited states
are collective modes of many Ho ions and should be classified in
momentum space. {\bf D} Calculated ratio $E_c/\Delta$ of the
minimum excitation energy $E_c$ to the single ion splitting
$\Delta$ at the critical field as a function of temperature. This
measures how far the electronic system is from the coherent limit,
for which $E_c/\Delta=0$. } \label{fig:gap}
\end{figure}

We investigate the excitation spectrum around the QPT in
LiHoF$_4$, which is an excellent physical realisation of the
transverse field Ising model, with an added term accounting for
the hyperfine coupling between electronic and nuclear moments
\cite{Wu,Bitko,BitkoII}. The dilution series
LiHo$_x$Y$_{1-x}$F$_4$ is the host for a wide variety of
collective quantum effects, ranging from tunnelling of single
moments and domain walls to quantum annealing, entanglement and
Rabi oscillations \cite{giraud,brooke99,brooke01,ghosh02,ghosh03}.
These intriguing properties rely largely on the ability of a
transverse field, whether applied externally or generated
internally by the off-diagonal part of the magnetic dipolar
interaction, to mix two degenerate crystal field states of each Ho
ion. The Ho ions in LiHoF$_4$ are placed on a tetragonal Scheelite
lattice with parameters $a=5.175$ \AA{} and $c=10.75$ \AA{}. The
crystal-field ground state is a $\Gamma_{3,4}$ doublet with only a
$c$ component to the angular momentum and hence can be represented
by the $\sigma^z=\pm1$ Ising states. A transverse field in the
$a$-$b$ plane mixes the higher-lying states with the ground state
producing a splitting of the doublet, equivalent to an effective
Ising model field. The phase diagram  of LiHoF$_4$ (Fig.\
\ref{fig:gap}A) was determined earlier by susceptibility
measurements \cite{Bitko}, and displays a zero field $T_c$ of
1.53~K, and a critical field of $H_c=49.5$~kOe in the zero
temperature limit. The same measurements confirmed the strong
Ising anisotropy, with longitudinal and transverse $g$-factors
differing by a ratio of 18 \cite{Bitko}. The sudden increase in
$H_c$ below 400~mK was explained by alignment of the Ho nuclear
moments through the hyperfine coupling. Corrections to phase
diagrams due to hyperfine couplings have a long history
\cite{Andres}, and were noted for the Li{\it RE\/}F$_4$ ({\it
RE}=Rare earth) series, of which LiHoF$_4$ is a member, over
twenty years ago \cite{youngblood}. What is new here is that
applying a transverse field and employing high resolution neutron
scattering spectroscopy allows us to carefully study the dynamics
as we tune through the quantum critical point.

\begin{figure}[h]
\includegraphics[width=0.85\linewidth]{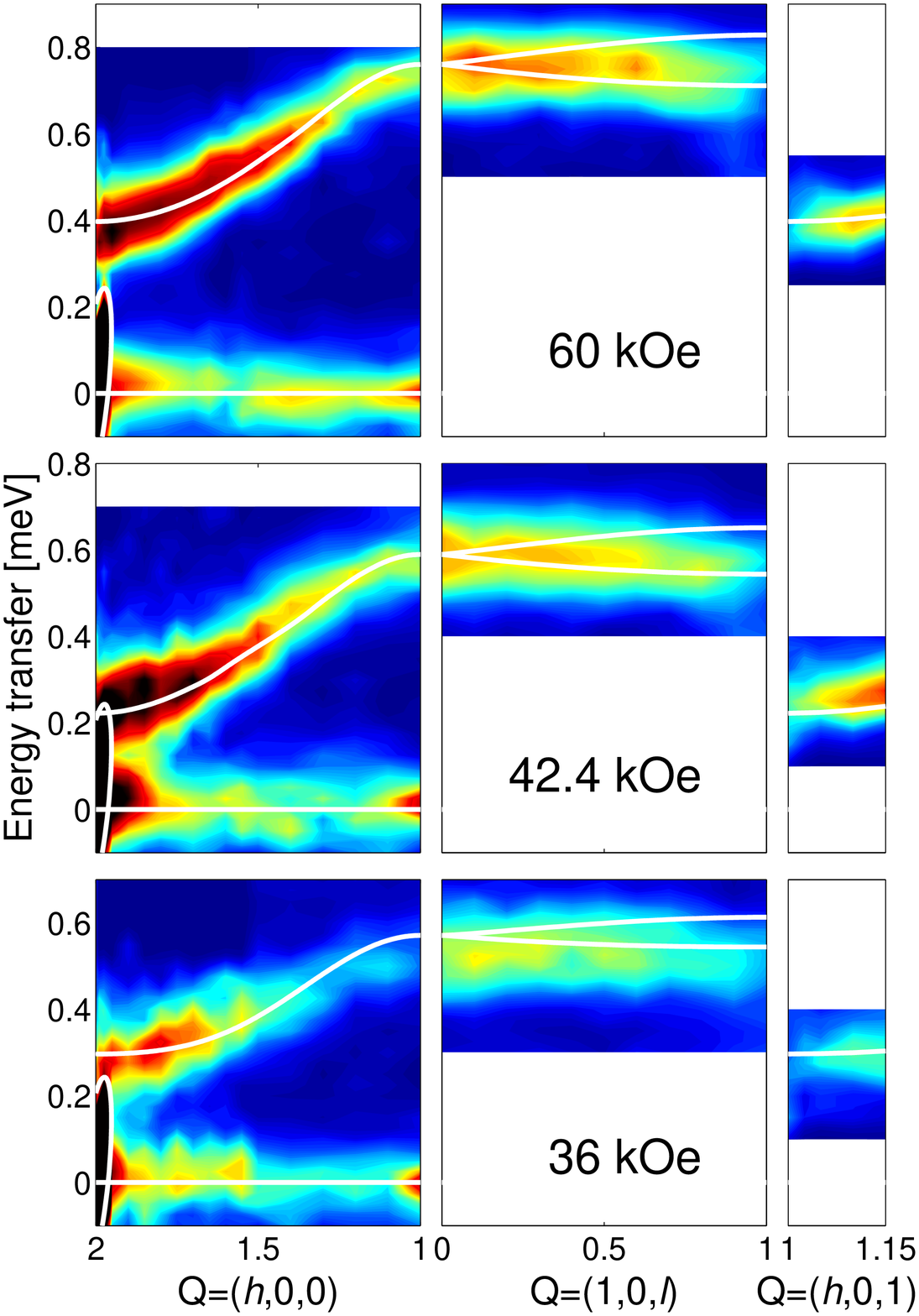}
\caption{Pseudocolor representation of the inelastic neutron
scattering intensity for LiHoF$_4$ at $T=0.31$~K observed along
the reciprocal space trace
$(2,0,0)\rightarrow(1,0,0)\rightarrow(1,0,1)\rightarrow(1.15,0,1)$.
White lines show the $1/z$ calculation for the excitation energies
as described in the text. White ellipses around the (2,0,0) Bragg
peak indicate 5 times the full-width-half-max resolution tail.}
\label{fig:map}
\end{figure}

We measured the magnetic excitation spectrum of LiHoF$_4$ using
the TAS7 neutron spectrometer at Ris\o{} National Laboratory, with
an energy resolution (full width at half maximum) of
0.06-0.18~meV\cite{Henrik}. The transverse field was aligned to
better than $0.35^\circ$ and the sample cooled in a dilution
refrigerator. At the base temperature of 0.31~K, giving a critical
field of 42.4~kOe, the excitation spectrum was mapped out below,
at and above the critical field (Fig. \ref{fig:map}). For all
fields a single excitation branch disperses upwards from a minimum
gap at (2,0,0) towards (1,0,0). From (1,0,0) to (1,0,1), the mode
shows little dispersion but appears to broaden. The discontinuity
on approaching respectively $(1,0,1-\epsilon)$ and
$(1+\epsilon,0,1)$ reflects the anisotropy and long-range nature
of the magnetic dipole coupling. However, the most important
observation is that the (2,0,0) energy, which is always lower than
the calculated single ion energy ($\sim 0.39$ meV at 42.4~kOe),
shrinks on increasing the field from 36 to 42.4~kOe, and then
hardens again at 60~kOe. At this qualitative level, what we see
agrees with the mode softening predicted for the simple Ising
model in a transverse field. However, it appears that the mode
softening is incomplete. At the critical field of 42.4~kOe it
retains a finite energy of $0.24\pm0.01$~meV. Fig. \ref{fig:gap}B,
showing the gap energy as a function of the external field, makes
this result very apparent.

To obtain a quantitative understanding of our experiments we
consider the full rare-earth Hamiltonian, which closely resembles
that of HoF$_3$ \cite{HoF3a,HoF3b}. Each Ho-ion is subject to the
crystal field, the Zeeman and the hyperfine coupling. The
interaction between moments is dominated by the long-range dipole
coupling, with a small nearest neighbour exchange interaction
${\mathcal J}_{12}^{}$:
\begin{eqnarray} {\mathcal H}&=&\sum_i\Big[{\mathcal
H}_{\mathrm{CF}}^{}({\bf J}_i^{})+ A\, {\bf J}_i^{}\cdot{\bf
I}_i^{}-g\mu_B^{}{\bf J}_i^{}\cdot{\bf H}\Big]\nonumber
\\&-&{\textstyle\frac{1}{2}}\sum_{ij}
\sum_{\alpha\beta}{\mathcal
J}_D^{}D_{\alpha\beta}^{}(ij)J_{i\alpha}^{}J_{j\beta}^{}
-{\textstyle\frac{1}{2}}\sum_{ij}^{n.n.}{\mathcal J}_{12}^{}\,{\bf
J}_i\cdot{\bf J}_j^{} \quad\label{2}
\end{eqnarray}
\noindent where $\mathbf{J}$ and $\mathbf{I}$ are respectively the
electronic and nuclear moments, and for $^{165}$Ho$^{3+}$ $J=8$
and $I=7/2$. Hyperfine resonance \cite{Magarino} and heat capacity
measurements \cite{Mennenga} show that the hyperfine coupling
parameter $A=3.36~\mu$eV as for the isolated ion, with negligible
nuclear-quadrupole coupling. The Zeeman term is reduced by the
demagnetization field. The normalized dipole tensor
$D_{\alpha\beta}(ij)$ is directly calculable, and the dipole
coupling strength is simply fixed by lattice constants and the
magnetic moments of the ions at ${\mathcal
J}_D=(g\mu_B)^2N=1.1654~\mu$eV. This leaves as free parameters
various numbers appearing in the crystal field Hamiltonian
${\mathcal H}_{\mathrm{CF}}$ and the exchange constant ${\mathcal
J}_{12}^{}$. The former are determined\cite{JJnew} largely from
electron spin resonance for dilute Ho atoms substituted for Y in
LiYF$_4$, while the latter is constrained by the phase diagram
determined earlier \cite{Bitko} (Fig.1 A). We have used an
effective medium theory \cite{stinchcombe} previously applied to
HoF$_3$ \cite{JJz} to fit the phase diagram, and conclude that a
good overall description, except for a modest (14\%) overestimate
of the zero field transition temperature is obtained for
${\mathcal J}_{12}^{}=-0.1~\mu$eV. Based on quantum Monte Carlo
simulation data others \cite{girvin} have also concluded that
${\mathcal J}_{12}^{}$ is substantially smaller than ${\mathcal
J}_{D}^{}$.

Having established a good parametrization of the Hamiltonian, we
model the dynamics, where expansion to order $1/z$ ($z$ is the
number of nearest neighbors of an ion in the lattice) leads to an
energy dependent renormalization $[1+\Sigma(\omega)]^{-1}$ (of the
order 10\%) of the dynamic susceptibility calculated in the random
phase approximation (RPA), with $\Sigma(\omega)$ evaluated as
described in \cite{JJz}. For the three fields investigated in
detail, the dispersion measured by neutron scattering is closely
reproduced throughout the Brillouin zone. As indicated by the
solid lines in Fig.\ \ref{fig:map}, the agreement becomes
excellent if the calculated excitation energies are multiplied by
a renormalization factor $Z=1.15$. The point is not that the
calculation is imperfect, but rather that it matches the data as
closely as it does. Indeed, it also predicts a weak mode splitting
of about 0.08~meV at $(1,0,1-\epsilon)$, consistent with the
increased width in the measurements. The agreement for the
discontinuous jump between $(1,0,1-\epsilon)$ and
$(1+\epsilon,0,1)$ due to the long-range nature of the dipole
coupling, shows that this is indeed the dominant coupling.

\begin{figure}[t]
\includegraphics[width=0.85\linewidth]{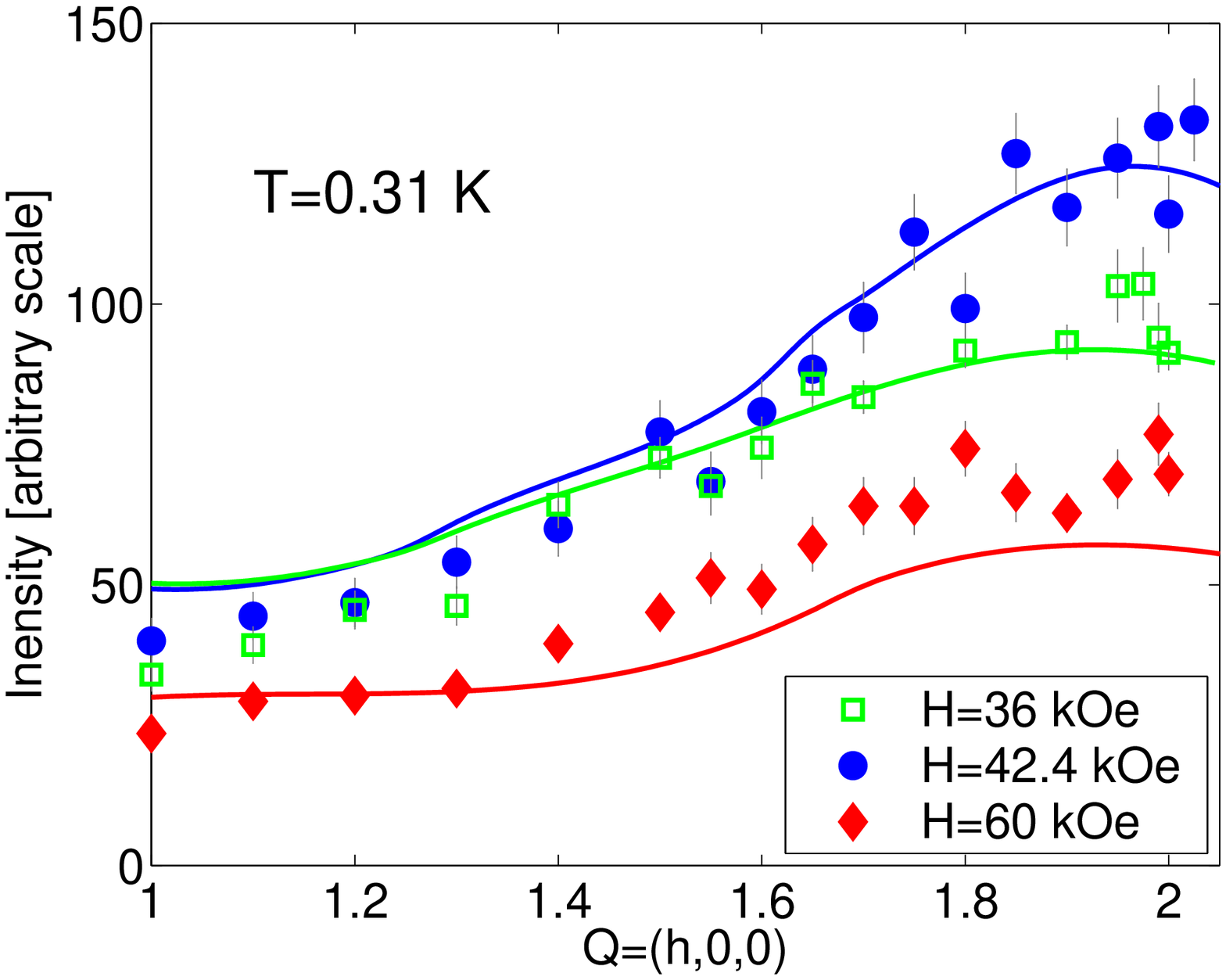}
\caption{Measured intensities of the excitations along $Q=(h,0,0)$
at the same values of the field as in Fig.\ 2. Lines are
calculated with geometric and resolution corrections applied to
allow comparison to the neutron data.} \label{fig:int}
\end{figure}

Fig.\ \ref{fig:gap}B-C illustrates the simple origin of the
incomplete softening and enhanced critical field, which is easiest
to understand if we start from the polarised paramagnetic state
above $H_c$, where the experiment, the purely electronic
calculation, and the theory including the hyperfine coupling all
coincide. At high fields the only effect of the hyperfine term is
to split both the ground state and the electronic excitation modes
into multiplets which are simply the direct products of the
electronic and nuclear levels, with a total span of $2A\langle
J\rangle I\simeq0.1$~meV as illustrated in Fig.\ \ref{fig:gap}C.
Upon lowering the field, the electronic mode softens and would in
the absence of hyperfine coupling reach zero energy at
$H_c^0=36$~kOe. The hyperfine coupling, however, mixes the
original ground and excited (soft mode) states already above
$H_c$. As this happens, the formation of a composite spin from
mixed nuclear and electronic contributions immediately stabilizes
ordering along the $c$-axis of the crystal. In other words, the
hyperfine coupling shunts the electronic mode, raising the
critical field to the observed $H_c$=42.4~kOe, where the mode
reaches a non-zero minimum. This process is accompanied by
transfer of intensity from the magnetic excitation of electronic
origin to much lower energy (in the $10~\mu$eV range) soft modes
of entangled nuclear/electronic character. Cooling to very low
temperatures would reveal these modes as propagating and softening
to zero at the quantum critical point, but at the temperatures
reachable in our measurements there is thermalization, dephasing
the composite modes to yield the strong quasi-elastic scattering
appearing around $Q=(2,0,0)$ and zero energy at the critical
field, as in Fig. \ref{fig:map}.

The intensities of the excitations are simply proportional to the
matrix elements $|\langle f|\sum_j\exp(iQ\cdot
R_j)J_j^+|0\rangle|^2$, and therefore provide a direct measure of
the wavefunctions via the interference effects implicit in the
spatial Fourier transform of $J_j^{}$. Fig.\ \ref{fig:int} shows
intensities recorded along $(h,0,0)$ for the three fields 36,
42.4, and 60~kOe. They follow a momentum dependence characterized
by a broad peak near (2,0,0), which is well described by our
theory. In the absence of hyperfine interactions, the intensity at
$H_c^0$ would diverge as $q$ approaches $(2,0,0)$, reflecting that
the real-space dynamical coherence length $\xi_c$ of the excited
state grows to infinity. The finite width of the peak observed at
$H_c$, corresponds in real space to a distance of order the
inter-holmium spacing, and implies that because they forestall the
softening of the electronic mode, the hyperfine interactions also
limit the distance over which the electronic wavefunctions can be
entangled\cite{osterloh02}. Thus, Fig.\ \ref{fig:int} is a direct
demonstration of the limitation of quantum coherence in space via
coupling to a nuclear spin bath. The dynamical length $\xi_c$ is
obtained from a sum over matrix elements connecting the ground
state to a particular set of excited states, while the
thermodynamic correlation length $\xi_t$ is derived from the equal
time correlation function $S(r)$ which is the sum over all final
states. $\xi_t$ diverges at second order transitions such as those
in LiHoF$_4$, where the quasielastic component seen in our data
dominates the long-distance behaviour of $S(r)$ at $T_c(H)$. It is
the electronic mode and hence $\xi_c$ that dictates to which
extent LiHoF$_4$ can be characterised and potentially exploited as
a realisation of the ideal transverse field Ising model.

Beyond providing a quantitative understanding of the excitations
near the quantum critical point of a model experimental system, we
obtain new insight by bringing together the older knowledge from
rare-earth magnetism and the contemporary ideas of entanglement,
qubits and decoherence. While the notion of the spin-bath was
developed to address decoherence in localised magnetic clusters
and molecules\cite{stamp1}, our work discloses its significance
for quantum phase transitions. Notably, we establish that the spin
bath is a generic feature that will limit how far it is possible
to observe intrinsic electronic quantum criticality. While this
may not be a significant factor when thinking about transition
metal oxides with very large exchange constants, it could matter
for rare earth and actinide intermetallic compounds which show
currently unexplained cross-overs to novel behaviours at low
($<1$~K) temperatures (see e.g. \cite{gegenwart02}).

For magnetic clusters, decoherence can be minimised in a window
between the oscillator-bath dominated high temperature and the
spin-bath dominated low-temperature regions\cite{stamp2}. Our
calculations suggest that the dense quantum critical magnet shows
analogous behaviour. Here the interacting electron spins
themselves constitute the oscillator bath, and the extent to which
the magnetic excitation softens at $T_c(H)$, measured by the ratio
of the zone center energy $E_c$ to the field-induced single ion
splitting $\Delta$ (Fig.\ \ref{fig:gap}D), gauges the electronic
decoherence. $E_c/\Delta$ achieves its minimum not at $T=0$, but
rather at an intermediate temperature $T\simeq1$~K, exactly where
the phase boundary in Fig.\ \ref{fig:gap}A begins to be affected
by the nuclear hyperfine interactions.

\bibliography{lihof4}
\bibliographystyle{Science}

\begin{scilastnote}
\item We thank  G. McIntyre for his expert assistance during
complementary measurements on the D10 diffractometer at the
Institut Laue Langevin, Grenoble, France. Work at the University
of Chicago was supported primarily by the NSF MRSEC Program under
DMR-0213745. Work in London was supported by the Wolfson-Royal
Society Research Merit Award Program and the Basic Technologies
program of the UK Research Councils.
\end{scilastnote}

\end{document}